\documentclass[twocolumn,preprintnumbers,prb,amsmath,amssymb]{revtex4-2}

\usepackage{graphicx}
\usepackage{dcolumn}
\usepackage{bm}
\usepackage{color}
\usepackage{mathrsfs}

\begin{document}

\preprint{}

\title{ 
Scaling theory of charge transport and thermoelectric response in disordered 2D electron systems: From weak to strong localization}
\author{Takahiro Yamamoto$^{1,2}$}
\author{Hiroki Kaya$^1$}
\author{Manaho Matsubara$^{1,2}$}
\author{Hidetoshi Fukuyama$^2$}

\affiliation{$^1$Department of Physics, Tokyo University of Science, Kagurazaka 1-3 Shinjuku, Tokyo 162-8601, Japan}
\affiliation{$^2$RIST, Tokyo University of Science, Kagurazaka 1-3 Shinjuku, Tokyo 162-8601, Japan}



\begin{abstract} 
We develop a new theoretical scheme for charge transport and thermoelectric response in two-dimensional disordered systems exhibiting 
crossover from weak localization (WL) to strong localization (SL). The scheme is based on the scaling theory for Anderson localization combined with the Kubo--Luttinger theory.
Key aspects of the scheme include introducing a unified  $\beta$ function that seamlessly connects the WL and SL regimes, 
as well as describing the temperature ($T$) dependence of the conductance from high to low $T$ regions on the basis of the dephasing length.
We found that the Seebeck coefficient, $S$, behaves as $S\propto T$ in the WL limit and as $S\propto T^{1-p}$ ($p < 1$) in the SL limit,
both with possible logarithmic corrections. The scheme is applied to analyze experimental data for thin films of the p-type organic semiconductor 
poly[2,5-bis(3-alkylthiophen-2-yl)thieno(3,2-b)thiophene] (PBTTT).
\end{abstract}

\maketitle 
\section{Introduction}
The development of high-performance thermoelectric (TE) materials is a focus of extensive research, particularly in relation to sustainable energy production. 
Since Hicks and Dresselhaus suggested in 1993 that a substantial increase in TE performance could be achieved by using low-dimensional 
semiconductors~\cite{rf:hicks}, numerous low-dimensional  TE materials have been discovered. 
Recently, poly[2,5-bis(3-alkylthiophen-2-yl)thieno(3,2-b)thiophene] (PBTTT) thin films, composed of $\pi$-stacked PBTTT polymers, have garnered significant attention as a new class of two-dimensional (2D) p-type organic semiconductors, both experimentally~\cite{rf:wang,rf:Venkateshvaran,rf:glaudell,rf:tanaka,rf:Fratini,rf:ito,rf:watanabe,rf:chen,rf:ito_private,rf:kang,rf:zanettini,rf:yamashita,rf:sd-kang,rf:huang} and  theoretically~\cite{rf:Northrup,rf:li}, due to their high TE performance.
Because of the inherent structural disorder of polymers, understanding and controlling their charge transport and TE response pose important challenges 
in condensed matter physics and materials engineering. 

Interesting experimental results on charge transport in p-type PBTTT thin film were reported by several groups~\cite{rf:tanaka,rf:ito, rf:sd-kang, rf:huang, rf:ito_private}.
Recently, Ito {\it et al.} measured the temperature ($T$) dependence of electrical conductivity, $\sigma(T)$, for various hole densities controlled by 
an electrochemical transistor (ECT)~\cite{rf:ito_private}. They found that $\sigma(T)$ exhibits a logarithmic $T$ dependence, characteristic of weak localization (WL),
in the range 200--100K. As $T$ decreases from 100 to 30K, $\sigma(T)$ deviates from this logarithmic behavior, which was attributed to a precursor to the crossover 
from WL to strong localization (SL), characterized by an exponential $T$ dependence of the conductivity. Similarly, experimental data on $\sigma(T)$ for similar hole 
density achieved by chemical doping have been reported by Watanabe {\it et al.}~\cite{rf:watanabe}, who analyzed the data under the assumption of 
variable range hopping (VRH)-type exponential $T$ dependence.

In addition to electrical conductivity, the Seebeck coefficient, $S(T)$, has also been measured in a wide $T$ range by Ito {\it et al.}~\cite{rf:ito_private}
and Watanabe {\it et al.}~\cite{rf:watanabe}. 
Despite the differences in carrier-doping methods, both groups conclude that 
$S(T)$ is proportional to $T$ at high carrier densities, apparently consistent with the Mott formula for metals~\cite{rf:mott-davis}, whereas 
$S(T)$ deviates from $T$-linear behavior at low carrier densities. A coherent understanding of this behavior has not yet been achieved.

The purpose of the present study is to theoretically elucidate, on the basis of the scaling theory of Anderson localization, 
 the $T$ dependence of $\sigma(T)$ and $S(T)$ in the WL-SL crossover regime in 2D systems.
This approach aims to provide a unified understanding of various experimental data 
obtained through different carrier-doping methods~\cite{rf:ito_private, rf:watanabe}, for which a consensus on the physical interpretation has not been attained.

\section{Theoretical scheme}
\subsection{Scaling theory at finite temperature}
The scaling theory for the Anderson localization~\cite{rf:nagaoka}, {\it i.e.,} Abraham--Anderson--Licciardello--Ramakrishnan (AALR) theory~\cite{rf:aalr}, 
indicates that the conductance in a disordered system 
at $T=0$ follows a differential equation:
\begin{eqnarray}
\dfrac{d \ln g}{d \ln L}=\beta(g),
\label{eq:RG-eq}
\end{eqnarray}
where $g$ is the dimensionless conductance scaled by the universal conductance $e^2/h$ ($e$: elementary charge, $h$: Planck's constant), 
$L$ is the length of the system, and $\beta(g)$ is assumed to be a smoothly and monotonically increasing function with respect to $g$.

In the large-$g$ limit, $g$ obeys the classical Ohm's law as $g=\sigma L^{d-2}$, where $\sigma$ is conductivity and $d$ is the spatial dimension of the system. 
From Eq.~(\ref{eq:RG-eq}), we see that the $\beta$ function in this limit is given by $\beta(g)= d-2$; the asymptotic behavior of $\beta(g)$ is therefore
expressed as $\beta(g)\approx d-2-\alpha/g$, where the expansion coefficient $\alpha$ is a positive constant on the order of unity.  
On the other hand, in the small-$g$ limit, the electronic states are localized because of disorder and $g$ decreases  exponentially with $L$ as $g\sim e^{-L/\xi}$, 
where $\xi$ is the localization length, resulting in $\beta(g)\approx \ln g$ in this limit. 
Thus, in the two limits, the $\beta$ function is given by
\begin{eqnarray}
\beta(g)\approx\left\{
\begin{array}{ll}
\ln g & (g\to 0) \\
d-2-\dfrac{\alpha}{g} & (g\to\infty)\\
\end{array}
\right.
\label{eq:asymptotic}
\end{eqnarray}
for $d$-dimensional systems. 
In three-dimensional (3D) systems ($d=3$), where conducting and localized states are sharply separated across the mobility edge,
different theoretical schemes are needed for each state. By contrast, the 2D systems ($d=2$) of present interest exhibit $g(L)\to 0$ in the limit of $L$; 
that is, the electronic states are always localized. If $L$ is finite and varied, $g(L)$ is expected to crossover from metallic to insulating behaviors as 
$L$ is increased. Actually, this expectation has been confirmed through studies on the frequency ($\omega$) dependences of conductance 
for $L=\infty$ at $T=0$ by Vollhardt and Wolfle~\cite{rf:Vollhardt-Wolfle} and by Kawabata~\cite{rf:kawabata} using the self-consistent (SC) theory.  
Notably, the SC theory is governed by a single parameter, $\lambda=\frac{\hbar}{\varepsilon_{\rm F}\tau}$
(where $\varepsilon_{\rm F}$ is the Fermi energy and $\tau$ is the elastic scattering time), which characterizes the scattering strength in quantum transport, 
and turns out to be capable of describing not only metallic but also insulating states. Regarding the $T$ dependence of conductance, however, 
we need to develop a different theoretical scheme. Once at finite temperatures, $g(T)$ can be non zero even for $L=\infty$ and the $T$ dependence of $g(T)$
exhibits a crossover from essentially metallic WL with $\ln T$ to SL with exponential $T$ dependences. The characteristic temperature of 
this crossover depends on the carrier density.

To address the WL--SL crossover in 2D systems, we first consider $T=0$ and introduce the following $\beta$ function as a smoothly and 
monotonically increasing function that asymptotically behaves as in Eq.~(\ref{eq:asymptotic}) in both limits of $g\to 0$ and $g\to \infty$:
\begin{eqnarray}
\beta(g)=-\ln\left(
\frac{\alpha}{g}+1
\right).
\label{eq:now-beta}
\end{eqnarray}
To solve the differential equation in Eq.~(\ref{eq:RG-eq}) with $\beta(g)$ in Eq.~(\ref{eq:now-beta}), 
we impose $g(L_0)=g_0$ as the boundary condition, where $L_0$ is the microscopic characteristic length that is much larger than 
the mean free path and $g_0$ is the conductance at length scale $L_0$. 
Under this condition, Eq.~(\ref{eq:RG-eq}) can be rewritten as
\begin{eqnarray}
\int_{g_0}^{g(L)}\dfrac{d\ln g}{\beta(g)}=\int_{L_0}^{L}d\ln L.
\label{eq:integral}
\end{eqnarray}
By solving Eq.~(\ref{eq:integral}) with Eq.~(\ref{eq:now-beta}), we obtain the $L$ dependence of $g$ at $T=0$.

To extend this theory to finite $T$, Anderson, Abrahams, and Ramakrishnan (AAR) proposed replacing the system length $L$ in Eq.~(\ref{eq:integral}) 
with a characteristic length $L_{\phi}$ (hereafter referred to as the {\it dephasing length}) over which an electron remains in an eigenstate~\cite{rf:aar} . 
In the diffusive transport in the WL regime, $L_{\phi}$ is given by $L_{\phi}=\sqrt{D\tau_{\phi}}$ with diffusion coefficient $D$ and dephasing time $\tau_{\phi}$. 
Because the dephasing time generally depends on $T$ as $\tau_{\phi}\propto T^{-2p}$ with a positive constant $p$, the $T$-dependence of  
$L_\phi$ is given by $L_{\phi}\propto T^{-p}$. In the present work, we adopt $L_\phi$ as a single parameter to describe the $T$ dependences, similar to
$\lambda=\frac{\hbar}{\varepsilon_{\rm F}\tau}$ in $L=\infty$ at $T=0$ in the SC theory, and express it as
\begin{eqnarray}
L_{\phi}(\varepsilon,T)=L_{\rm d}\left(\frac{T}{T_{\rm d}}\right)^{-p}
\label{eq:L_epsilon}
\end{eqnarray}
in terms of three different parameters: the temperature $T_{\rm d}$ characterizing the WL-SL crossover and the dephasing length $L_{\rm d}$ at $T = T_{\rm d}$
and the exponent $p$ of the temperature dependence.
This expression is possible because the $T$ dependence of conductance in a 2D system, unlike that in a 3D system with the mobility edge, exhibits a gradual change 
from logarithmic to exponential behavior in the $\omega$ dependences of the conductance as $T$ decreases.

In the WL limit, where $g_0$ is large and $g_0\gg\alpha\ln(L_\phi/L_0)=\alpha p\ln(T_{\rm WL}/T)$ is satisfied, 
the differential equation in Eq.~(\ref{eq:integral}), with $L_\phi(\varepsilon,T)$ substituted for $L$, 
can be solved  analytically. The spectral conductance $g(\varepsilon,T)$ is given by
\begin{eqnarray}
g(\varepsilon,T)&=&g_0(\varepsilon)-\alpha \ln\left(\frac{L_\phi(\varepsilon,T)}{L_0}\right)\nonumber\\
&\equiv&g_0(\varepsilon)-\alpha p(\varepsilon) \ln\left(\frac{T_{\rm WL}(\varepsilon)}{T}\right)
\label{eq:g_WL}
\end{eqnarray}
with 
\begin{eqnarray}
T_{\rm WL}(\varepsilon)\equiv  T_{\rm d}\left(\frac{L_{\rm d}(\varepsilon)}{L_0}\right)^{1/p(\varepsilon)},
\label{eq:T1}
\end{eqnarray}
resulting in the well-known logarithmic correction to conductance as long as $T< T_{\rm WL}(\varepsilon)$.
On the other hand, in the SL limit where $g_0$ is small ($\alpha/g_0\gg 1$), the differential equation can also be analytically solved, 
enabling $g(\varepsilon,T)$ to be expressed as
\begin{eqnarray}
g(\varepsilon,T)=\alpha\left(\frac{g_0}{\alpha}\right)^{\frac{L_\phi}{L_0}}
\equiv\alpha \exp\left[-\left(\frac{T_{\rm SL}(\varepsilon)}{T}\right)^{p(\varepsilon)}\right]
\label{eq:g_SL}
\end{eqnarray}
for $T\ll T_{\rm SL}(\varepsilon)$, where $T_{\rm SL}(\varepsilon)$ is defined as 
\begin{eqnarray}
T_{\rm SL}(\varepsilon)\equiv  T_{\rm d} \left(\frac{L_{\rm d}(\varepsilon)}{L_0}\right)^{1/p(\varepsilon)}\left(\ln\frac{\alpha}{g_0(\varepsilon)}\right)^{1/p(\varepsilon)}.
\label{eq:T0}
\end{eqnarray}

\subsection{Thermoelectric response in WL-SL crossover}
The electrical conductivity and the Seebeck coefficient can be expressed in terms of $g(\varepsilon, T)$ as follows.
In linear response theory, the electrical current density ${\bm j}$ under the electric field ${\bm E}$ and the temperature gradient $\nabla T$
can be described by
$\displaystyle {\bm j}=L_{11}{\bm E}-L_{12}({\nabla T}/{T})$.
Here, the electrical conductivity $L_{11}(=\sigma)$ and the thermoelectric conductivity $L_{12}$ can be expressed by the following Sommerfeld-Bethe (SB) relation
based on the Kubo--Luttinger theory~\cite{rf:kubo,rf:Luttinger} under the assumption that the heat current is carried only by electrons.
\begin{eqnarray}
L_{11}(T) &=& \frac{2e^2}{h}\frac{L}{A_{\rm c}}\int_{-\infty}^\infty \!\!\! d\varepsilon \left( -\frac{\partial f_0}{\partial \varepsilon} \right)g(\varepsilon,T),
\label{sigmaxx_SL}\\
L_{12}(T) &=& -\frac{2e}{h}\frac{L}{A_{\rm c}}\int_{-\infty}^\infty \!\!\! d\varepsilon \left( -\frac{\partial f_0}{\partial \varepsilon} \right) 
(\varepsilon-\mu) g(\varepsilon,T),
 \label{SB_relation_SL}
\end{eqnarray}
where the factor of 2 accounts for the spin degree of freedom, $L$ and $A_{\rm c}$ represent the length and cross-sectional area of the system, respectively,
$f_0$ denotes the Fermi--Dirac distribution function, and $\mu$ is the chemical potential. Notably, the SB relation was originally derived from 
the Boltzmann transport theory (BTT); however, it is valid even for strongly disordered systems that cannot be treated by the BTT~\cite{rf:MO_HF}.
Using Eq.~(\ref{sigmaxx_SL}) and Eq.~(\ref{SB_relation_SL}), we can describe the Seebeck coefficient by
\begin{eqnarray}
S=\dfrac{1}{T}\dfrac{L_{12}(T)}{L_{11}(T)}
=\dfrac{-1}{eT}\dfrac{\int_{-\infty}^\infty\! d\varepsilon \left( -\frac{\partial f_0}{\partial \varepsilon} \right) 
(\varepsilon-\mu) g(\varepsilon,T)}{\int_{-\infty}^\infty\! d\varepsilon \left( -\frac{\partial f_0}{\partial \varepsilon} \right)g(\varepsilon,T)}.
\label{eq:seebeck_LR}
\end{eqnarray}
Here, we note that the $\varepsilon$ dependences of $g(\varepsilon, T)$ results from those of $g_0$, $p$, $T_{\rm WL}$ and 
$T_{\rm SL}$ in the present scheme. When the $\varepsilon$ dependence of $g(\varepsilon,T)$ within $k_{\rm B}T$ is weak and
can be approximated as linear with respect to $\varepsilon$, {\it i.e.}, $g(\varepsilon,T)\approx g(\mu,T)+g'(\mu,T)(\epsilon-\mu)$, 
the Seebeck coefficients in the two limits are respectively given as
\begin{eqnarray}
S_{\rm WL}\approx S_0^{\rm WL}(\mu)\frac{T}{T_{\rm WL}(\mu)}\left(
1+A(\mu)\ln\frac{T_{\rm WL}(\mu)}{T}
\right)
\label{eq:seebeck_WL_low-T}
\end{eqnarray}
and
\begin{eqnarray}
S_{\rm SL}&\approx& S_0^{\rm SL}(\mu)\left(\frac{T}{T_{\rm SL}(\mu)}\right)^{1-p(\mu)}\nonumber\\
&&\times \left(1+B(\mu)\ln\frac{T_{\rm SL}(\mu)}{T}\right).
\label{eq:seebeck_SL_low-T}
\end{eqnarray}
Here, $S_0^{\rm WL}(\mu)$, $A(\mu)$ and $S_0^{\rm SL}(\mu)$, $B(\mu)$ are respectively given by
\begin{eqnarray}
S_0^{\rm WL}(\mu)&=&-\frac{\pi^2 k_{\rm B}^2T_{\rm WL}(\mu)}{3e}\frac{g'_0(\mu)}{g_0(\mu)}\left(1-\Delta(\mu)\right),
\label{eq:A}\\
A(\mu)&=&\alpha\frac{\frac{p(\mu)}{g_0(\mu)}(1-\Delta(\mu))-\frac{p'(\mu)}{g'_0(\mu)}}{1-\Delta(\mu)}
\label{eq:B}
\end{eqnarray}
with $\Delta(\mu)=\alpha\frac{p(\mu)}{g'_0(\mu)}\frac{T'_{\rm WL}(\mu)}{T_{\rm WL}(\mu)}$, and
\begin{eqnarray}
S_0^{\rm SL}(\mu)&=&\frac{\pi^2 k_{\rm B}^2}{3e}p(\mu)T_{\rm SL}'(\mu),
\label{eq:C}\\
B(\mu)&=&\frac{p'(\mu)}{p(\mu)}\frac{T_{\rm SL}(\mu)}{T_{\rm SL}'(\mu)}.
\label{eq:D}
\end{eqnarray}
Thus, $S_{\rm WL}(T)$ is proportional to $T$, whereas $S_{\rm SL}(T)$ is proportional to $T^{1-p}$ and
both of which have logarithmic corrections.

\begin{figure}[t]
  \begin{center}
  \includegraphics[keepaspectratio=true,width=80mm]{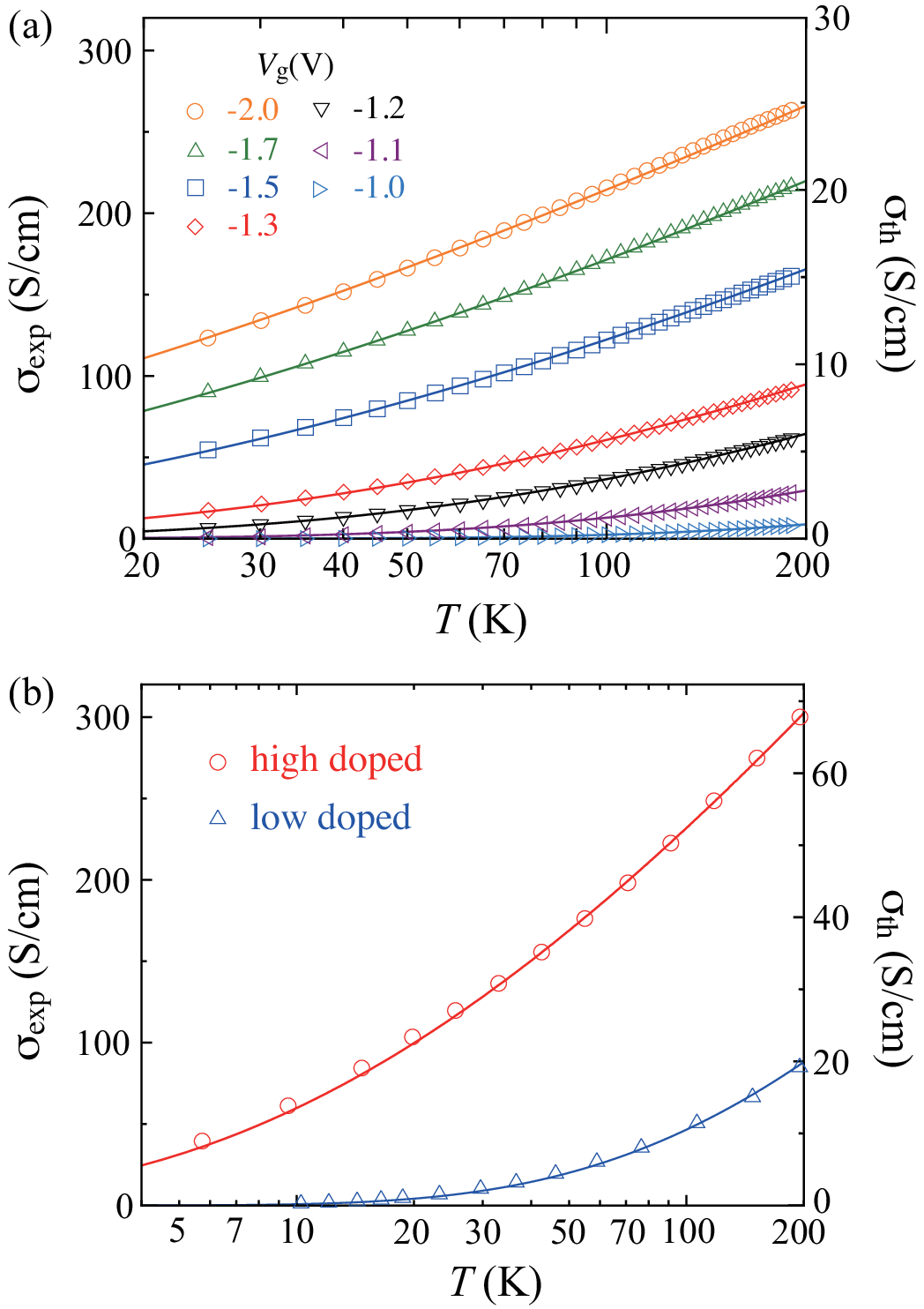}
  \end{center}
\caption{(Color online)
(a) $T$ dependence of the electrical conductivity of a PBTTT thin film under various gate voltages.
The marks are experimental data acquired by Ito {\it et al.} using the ECT~\cite{rf:ito_private}.
The solid curves are the present theoretical results.
The left and right vertical axes correspond to the experimental and theoretical results, $\sigma_{\rm exp}$ and $\sigma_{\rm th}$, respectively.
(b) $T$ dependence of the electrical conductivity of chemically carrier-doped PBTTT thin films.
The marks are experimental data reported by Watanabe {\it et al.}~\cite{rf:watanabe}. 
}
\label{fig:01}
\end{figure}

\section{Comparison with experiments}
\subsection{$T$ dependence of electrical conductivity}
We here compare the present theory to the experimental data for PBTTT thin films
obtained by two different carrier-doping methods: the electrochemical doping method~\cite{rf:ito_private} and the chemical doping method~\cite{rf:watanabe}.
For the ECT fabricated by Ito {\it et al.}~\cite{rf:ito_private}, the channel length between the source and drain electrodes is $L=140$~$\mu$m, 
the channel width is $W=2$~mm, and the film thickness is $t=17$~nm.
On the other hand, for the PBTTT device fabricated by Watanabe {\it et al.}~\cite{rf:watanabe}, $L=300$~$\mu$m, $W=80$~$\mu$m, and $t=40$~nm. 
Note that the lamellar spacing of PBTTT (a distance between two layers in the PBTTT with a multi layered 2D structure) is $d_l= 2$~nm; thus, 
the cross-sectional area of a single layer in Eqs.~(\ref{sigmaxx_SL}) and (\ref{SB_relation_SL}) is given by $A_{\rm c}\equiv Wd_l$. 

Figure~\ref{fig:01}(a) presents the experimental electrical conductivity, $\sigma_{\rm exp}(T)$, under various gate voltages, $V_g$, as measured using 
the ECT setup~\cite{rf:ito_private}. 
$\sigma_{\rm exp}(T)$ increases monotonically with increasing $|V_g|$ at a fixed $T$, which is reasonable because the hole density increases. 
We also observe that $\sigma_{\rm exp}(T)$ is proportional to $\ln T$ in the higher-$T$ region and deviates from this logarithmic behavior
as $T$ decreases, which we ascribe to a precursor to the WL--SL crossover. 
The solid curves in Fig.~\ref{fig:01}(a) represent the $T$ dependences of theoretical results $\sigma_{\rm th}(T)$ 
based on the assumption of $L_0=4$nm, $\alpha=1.0$, and $T_{\rm d}=20$K, where $T_d$ is chosen as a typical temperature
at which $\sigma_{\rm exp}(T)$ in Fig.~\ref{fig:01} deviates from the logarithmic behavior.
The three qualitatively different parameters ($g_0$, $L_{\rm d}$, and $p$) obtained by fitting to experimental data are listed in TABLE~\ref{tab:parameters},
showing convincing agreement, except for the difference between the absolute values of $\sigma_{\rm exp}$ and $\sigma_{\rm th}$. 
We consider that the difference is due to uncontrolled sample conditions, such as layer numbers and types of spatial disorder.

Figure~\ref{fig:01}(b) shows the electrical conductivity of PBTTT thin films chemically doped at two different doping levels. 
The marks represent experimental data~\cite{rf:watanabe} and the solid curves are the present theoretical results
based on the same assumption of $L_0=4$nm, $\alpha=1.0$, $T_{\rm d}=20$K. The values of the three parameters ($g_0$, $L_{\rm d}$ and $p$) are 
listed in TABLE~\ref{tab:parameters_2}. 
The theoretical curves are in excellent agreement with the experimental data, except for the difference between their absolute values, 
similar to Fig.~\ref{fig:01}(a).

\begin{table}[t]
  \caption{Parameters $g_0$, $L_d$ and $p$ as a function of $V_g$ for the PBTTT-based ECT in Ref.~\cite{rf:ito_private}. 
  Here, $L_0=4$nm, $\alpha=1.0$ and $T_{\rm d}=20$K are assumed. }
  \begin{ruledtabular}
    \begin{center}
      \begin{tabular}{cc|cccccc}
      $V_{g}~({\rm V})$ & & $g_0$ & & $L_{\rm d}~({\rm nm})$ & & $p$   & \\
      \hline  
       $-1.0$ & & 0.55   & &   49.5   & &  0.449  \\
       $-1.1$ & & 0.88   & &   54.0   & &  0.437  \\      
       $-1.2$ & & 1.24   & &   59.3   & &  0.422   \\
       $-1.3$ & & 1.51  & &    64.4    & &  0.410  \\       
       $-1.5$ & & 2.02  & &    73.9     & &  0.393  \\
       $-1.7$ & & 2.31  & &    77.3     & &   0.392   \\             
       $-2.0$ & & 2.55  & &    80.5     & &   0.391  \\
      \end{tabular}      
    \end{center}
   \label{tab:parameters}
  \end{ruledtabular}
  \end{table}
  
  \begin{table}[t]
  \caption{Parameters $g_0$, $L_d$, and $p$ for the chemically carrier-doped PBTTT thin films in Ref.~\cite{rf:watanabe}.  
  Here, $L_0=4$~nm, $\alpha=1.0$, and $T_{\rm d}=20$~K are assumed. }
    \begin{ruledtabular}
    \begin{center}
      \begin{tabular}{cc|cccccc}
      Doping level & & $g_0$ & & $L_d~({\rm nm})$ & & $p$   & \\
      \hline  
       low & & 0.92   & &   42.7   & &  0.441  \\
       high & & 2.08   & &   66.7   & &  0.383  \\      
      \end{tabular}      
    \end{center}
   \label{tab:parameters_2}
  \end{ruledtabular}  
\end{table}

As $V_g$ changes from $V_g=-1.0$~V to $-2.0$~V, $g_0$ increases monotonically from $0.55$ to $2.55$,
which reflects the carrier density and the strength of disorder scattering, as characterized by parameter $\lambda$ under the SC theory.
This result suggests that, under these experimental conditions, the electrons in the PBTTT thin film are situated more or less in the WL regime 
rather than in the SL regime. 
The data in TABLE~\ref{tab:parameters} also show that the dephasing length $L_{\rm d}$ increases and the exponent $p$ decreases 
because of the delocalization of electrons with increasing $|V_g|$.
Similar tendencies in $p$ and $L_d$ are observed in the data in TABLE~\ref{tab:parameters_2}.
Thus, the experimental data obtained by different carrier-doping methods can be understood in a unified manner using the three parameters 
$g_0$, $L_{\rm d}$ and $p$.

\begin{figure}[t]
  \begin{center}
  \includegraphics[keepaspectratio=true,width=80mm]{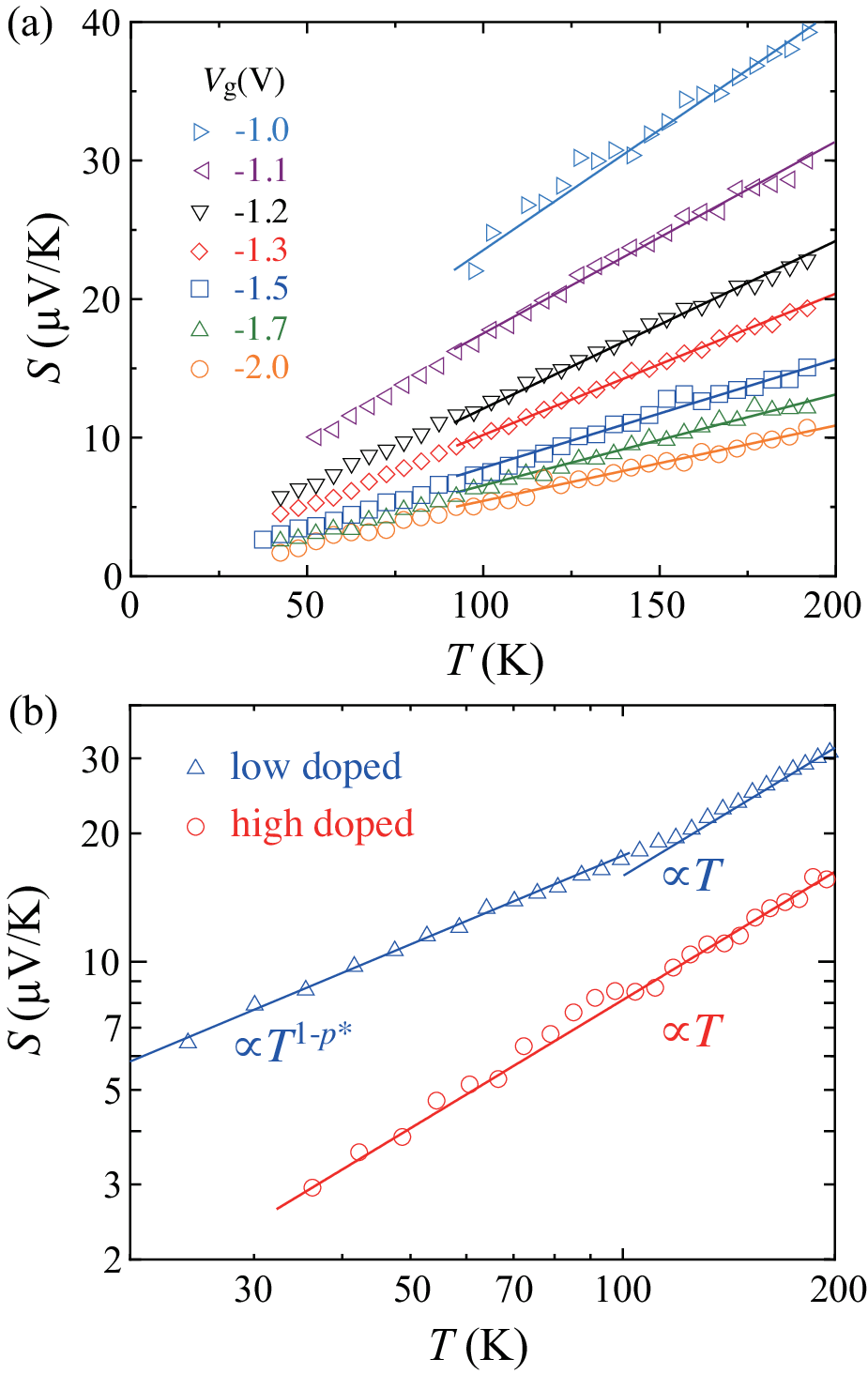}
  \end{center}
\caption{(Color online)
(a) $T$ dependence of Seebeck coefficient $S(T)$ of a PBTTT thin film under various gate voltages.
The marks are experimental data measured by Ito {\it et al.} using the ECT~\cite{rf:ito_private}, and the solid lines denote
the $T$-linear behavior in the WL regime in the high-$T$ region.
(b) $T$-dependence of $S(T)$ of chemically carrier-doped PBTTT thin films.
The marks are experimental data reported by Watanabe {\it et al.}~\cite{rf:watanabe}. The lines in the high-$T$ region represent
the $T$-linear behavior in the WL regime, and the line in the low-$T$ region denotes the power-law behavior as $S(T)\propto T^{1-p^*}$
with $p^*\sim 0.3$.
}
\label{fig:02}
\end{figure}

As shown in TABLE~\ref{tab:parameters} and TABLE~\ref{tab:parameters_2}, the exponent is $p\sim{0.4}$ which implies that $\tau^{-1}_\phi$ is 
almost proportional to $T$ in the crossover regime where diffusive motions are assumed in the microscopic region. 
This $T$ dependence might indicate that Coulomb interaction is the dominant scattering processes in this regime~\cite{rf:aalr2,rf:HF_Abrahams, rf:fukuyama}. 
It is of interest to see that the present result of $p\sim 1/2$ based on quantum transport is close to that argued by Efros and Shklovskii~\cite{rf:ES}, 
whose approach was based on the idea of VRH caused by electron--electron scattering.
Such an interesting possibility of interaction effects in the crossover regime between conducting and localized regimes is a very particular property of 2D systems,
in contrast to a sharp, though continuous, critical transition between them in 3D systems as analyzed recently for the Seebeck coefficient taking into account the energy 
dependences of the localization length near the mobility edge~\cite{rf:YOF}.
It is to be noted that $p=1/3$ in a 2D Mott VRH~\cite{rf:mott_SL}, where the electron hopping between localized states is caused by electron--phonon scattering.

\subsection{$T$ dependence of Seebeck coefficient}
We here discuss the $T$ dependence of $S(T)$ data for PBTTT thin films
obtained by the electrochemical doping method~\cite{rf:ito_private} and the chemical doping method~\cite{rf:watanabe}.
Figure~\ref{fig:02}(a) shows the experimental $S(T)$ of the PBTTT-based ECT under various $V_g$~\cite{rf:ito_private}.
The $S(T)$ for all $V_g$ are proportional to $T$ in the high-$T$ region, where the electrical conductivity varies as $\sigma(T)\propto \ln T$,
reflecting the WL. By fitting the experimental data using $S(T)=S_0^{\rm WL}\frac{T}{T_{\rm WL}}$ in Eq.~(\ref{eq:seebeck_WL_low-T}),
we determined $S_0^{\rm WL}$ and $T_{\rm WL}$, as listed in TABLE~\ref{tab:S_ECT}, where $T_{\rm WL}$ can be obtained from Eq.~(\ref{eq:T1}).
Both $S_0^{\rm WL}$ and $T_{\rm WL}$ increase as $V_g$ (or $g_0$) increases, as expected from Eqs.~(\ref{eq:T1}) and (\ref{eq:A}).
In the low-$T$ region for low-doped cases of $V_g=1.0$V and $1.1$V, $S(T)$ is expected to deviate from $T$-linear behavior; however, 
corresponding experimental data are not available.

Figure~\ref{fig:02}(b) shows the $S(T)$ for chemically carrier-doped PBTTT films~\cite{rf:watanabe}. In the case of high doping, 
$S(T)$ is proportional to $T$ within a wide $T$ region from 30 to 200~K, reflecting the WL. By contrast, in the case of low doping,
$S(T)$ is also linear with respect to $T$ in the high-$T$ region but deviates from the $T$-linear behavior in the low-$T$ region
(the fitting parameters $S_0^{\rm WL}$ and $T_{\rm WL}$ for the high- and low-doped cases are summarized in TABLE~\ref{tab:S_chemi-doped}).
In the low-$T$ region, $S(T)$ appears to vary as $S(T)\propto T^{1-p^*}$ with $p^*\sim 0.3$. 
Notably, the $T$-dependence of $S(T)$ follows a power-law behavior, which is a characteristic feature of SL as shown in Eq.~(\ref{eq:seebeck_SL_low-T}), 
even though the electrons in this low-doped PBTTT thin film are not in the fully SL regime but are somewhat more localized than in the WL regime.

\begin{table}[t]
  \caption{$S_0^{\rm WL}$ and $T_{\rm WL}$ for the PBTTT-based ECT used in Ref.~\cite{rf:ito_private}. 
Here, $L_0=4$~nm, $\alpha=1.0$ and $T_{\rm d}=20$~K are assumed. }
  \begin{ruledtabular}
    \begin{center}
      \begin{tabular}{cc|cccccc}
       $V_{\rm g}~({\rm V})$ & & $S_0^{\rm WL}$~(mV/K) & & $T_{\rm WL}$ (K) & & $S_0^{\rm WL}/T_{\rm WL}$~(nV/K$^2$) \\
      \hline     
       $-1.0$ & & $1.17$   & &   $5.42\times 10^{3}$  & & 215.8\\
       $-1.1$ & & $1.25$  & &    $7.72\times 10^{3}$  & & 161.9\\         
       $-1.2$ & & $1.44$   & &   $11.9\times 10^{3}$  & & 121.0\\
       $-1.3$ & & $1.80$  & &    $17.6\times 10^{3}$  & & 102.3\\       
       $-1.5$ & & $2.62$  & &    $33.5\times 10^{3}$  & &  78.3\\
       $-1.7$ & & $2.50$  & &    $38.1\times 10^{3}$  & &  65.6\\             
       $-2.0$ & & $2.36$  & &    $43.3\times 10^{3}$  & &  54.5\\
      \end{tabular}      
    \end{center}
   \label{tab:S_ECT}
  \end{ruledtabular}
\end{table}

\begin{table}[t]
  \caption{$S_0^{\rm WL}$ and $T_{\rm WL}$ for the chemically carrier-doped PBTTT thin films investigated in Ref.~\cite{rf:watanabe}. 
  Here, $L_0=4$~nm, $\alpha=1.0$, and $T_{\rm d}=20$~K are assumed. }
  \begin{ruledtabular}
    \begin{center}
      \begin{tabular}{cc|cccccc}
       doping level & & $S_0^{\rm WL}$~(mV/K) & & $T_{\rm WL}$ (K) & & $S_0^{\rm WL}/T_{\rm WL}$~(nV/K$^2$) \\
      \hline     
       low & & $0.70$   & &   $4.29\times 10^{3}$  & & 163.2\\
       high & & $2.61$  & &    $31.0\times 10^{3}$  & &  84.1\\         
      \end{tabular}      
    \end{center}
   \label{tab:S_chemi-doped}
  \end{ruledtabular}
\end{table}

\section{Summary and outlook}
We developed a new theoretical scheme for charge transport and TE response in disordered 2D electron systems.
The proposed scheme can describe the WL--SL crossover on the basis of the AALR+AAR theory~\cite{rf:aar, rf:aalr} combined with 
the Kubo--Luttinger theory~\cite{rf:kubo,rf:Luttinger}. 
The two key aspects of the scheme are (i) the introduction of the unified $\beta$ function, which seamlessly connects the WL and SL regimes, 
and (ii) the description of the $T$ dependence of the conductance from high and low $T$ regions in terms of $L_{\phi}(\varepsilon, T)$ in Eq.~(\ref{eq:L_epsilon}).
Using this scheme, we predicted $S(T)\propto T$ in WL and $S(T)\propto T^{1-p}$ in SL, both with logarithmic corrections. 

In addition, we applied this scheme to interpret recent experimental $\sigma(T)$ and $S(T)$ data for PBTTT thin films, which are realized by 
two different carrier doping methods~\cite{rf:ito_private,rf:watanabe}.
As a result, we could provide a unified theoretical interpretation for both experimental data based on the new scheme. We found that the electrons in 
the PBTTT thin films used in both experiments~\cite{rf:ito_private, rf:watanabe} are located more or less in the WL regime, not in the SL regime. We also found that $S(T)$ in 
PBTTT thin films exhibits $T$-linear behavior in the WL regime with a high carrier density but power-low behavior when the system begins to enter the SL regime from the WL regime as the carrier density decreases. 
Finally, we expect that the current theoretical scheme is applicable to other 2D materials (e.g., atomic layered materials) apart from present organic ones.

We thank Taishi Takenobu and Shun-ichiro Ito for fruitful discussions and for providing experimental data acquired using an ECT, 
We also thank Junichi Takeya and Shun Watanabe for valuable discussions regarding the experimental data obtained through the chemical doping method.
This work was partly supported by JSPS KAKENHI (Grant Nos. 22K18954 and 23H00259).

\end{document}